\documentclass[preprint,aps,pra]{revtex4}

\usepackage{graphicx}
\usepackage{amsfonts}
\usepackage{amssymb}
\usepackage{multirow}

\usepackage[normalem]{ulem}

\begin{document}

\title{Microscopic description of anisotropic low-density 
  dipolar Bose gases in two dimensions}

\author{A. Macia$^1$, F. Mazzanti$^1$, J. Boronat$^1$,
and R. E. Zillich$^2$}

\address{ $1$ Departament de F\'{\i}sica i Enginyeria Nuclear, Campus Nord
  B4-B5, Universitat Polit\`ecnica de Catalunya, E-08034 Barcelona, Spain}

\address{ $2$ Institut f\"ur Theoretische Physik, Johannes--Kepler
  Universit\"at, Altenbergerstr. 69, 4040 Linz, Austria}


\begin{abstract}

A microscopic description of the zero energy two-body ground state and
many-body static properties of anisotropic homogeneous gases of
bosonic dipoles in two dimensions at low densities is presented
and discussed. By changing the polarization angle with respect to the
plane, we study the impact of the anisotropy, present in the
dipole--dipole interaction, on the energy per particle, comparing the
results with mean field predictions. We restrict the analysis to the
regime where the interaction is always repulsive, although the
strength of the repulsion depends on the orientation with respect to
the polarization field. We present a series expansion of the solution
of the zero energy two-body problem which allows us to find the
scattering length of the interaction and to build a suitable Jastrow
factor that we use as a trial wave function for both a variational
and diffusion Monte Carlo simulation of the infinite
system.  We find that the anisotropy has an almost negligible impact
on the ground state properties of the many-body system in the
universal regime where the scattering length governs the physics of
the system. We also show that scaling in the gas parameter persists in
the dipolar case up to values where other isotropic interactions with
the same scattering length yield different predictions.

\end{abstract}


\maketitle

\narrowtext

\section{Introduction}

Quantum dipolar systems of bosons and fermions have gathered much
experimental and theoretical attention in recent years. In 2005,
Griesmaier and collaborators~\cite{Griesmaier_2005} on one side, and
Stuhler and collaborators~\cite{Stuhler_2005} on the other, reported
on the first experimental realization of a Bose condensate of
$^{52}$Cr, where the dipolar moment of the atoms is so large ($\sim
6\mu_B$) that the effect of the dipole--dipole interaction is
comparable in strength to the van der Waals forces.  More recently,
new and exciting results have been achieved with polar molecules of
Rubidium and Potassium ($^{40}$K$^{87}$Rb)~\cite{KKNi_2008}, which
have not been easy to create due to strong loss rates in the
population induced by chemical reactions~\cite{Ospelkaus_2009,
  KKNi_2010}.  A promising route towards a molecular Bose-Einstein
condensate is Feshbach association of Rb and Cs, which are not
reactive~\cite{Lercher_2011}.  One of the major advantages of polar
molecules is that the electric dipole moments are remarkably larger
than in the magnetic case of $^{52}$Cr and can be tuned by applying an
external electric field. Systems of polar molecules have been
speculated to present interesting applications ranging from the
control of chemical reactions~\cite{Krems_2008} to practical
applications of quantum information theory~\cite{Andre_2006}.

From the theoretical point of view, dipolar systems present novel and
interesting phenomena that make them particularly appealing. On the
one hand, the anisotropic character of the dipole-dipole interaction
introduces additional degrees of freedom compared with other condensed
matter systems that can potentially enrich the phase diagram. On the
other, the interaction decreases at large distances as $r^{-3}$ and
becomes long ranged in three dimensions (3D), in contrast to typical
van der Waals forces. In two dimensions (2D), though, the interaction
is still short ranged but at the border between both regimes.

\begin{figure}[b!]
\begin{center}
\includegraphics*[width=0.4\textwidth]{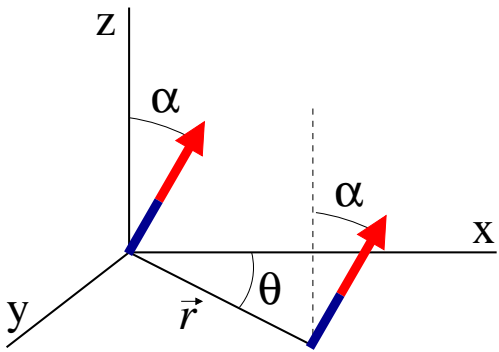}
\end{center}
\caption{Two dipoles confined to move on the X-Y
  plane. The polarization field lays on the XZ plane and 
  fixes a direction in space forming an angle $\alpha$ with the $z$ axis.}
\label{petardosdeferia}
\end{figure}

The potential $V_d(\bf r)$ describing how two dipoles with dipolar
moments ${\bf p}_1$ and ${\bf p}_2$ interact is given by
\begin{equation}
V_d({\bf r}) = {C_{dd} \over 4\pi}\,
\left[
{ \hat{\bf p}_1\cdot \hat{\bf p}_2 - 3(\hat{\bf p}_1\cdot\hat{\bf r}) 
(\hat{\bf p}_2\cdot\hat{\bf r}) \over r^3 }
\right]
\label{dipdipV}
\end{equation}
with ${\bf r}$ the relative position vector between them and
$C_{dd}$ the coupling constant defining the strength of the
interaction.  For permanent magnetic dipoles $C_{dd}=\mu_0 \mu^2$
where $\mu_0$ is the permeability of vacuum and $\mu$ is the permanent
dipole moment of the atoms.  Alternatively, the electric dipole moment
can be induced by an electric field ${\bf E}$, and in this case the
coupling constant is $C_{dd}=d^2/\epsilon_0$, where $d=\tilde\alpha E$ with
$\tilde\alpha$ the static polarizability and $\epsilon_0$ the permitivity
of vacuum.  For a system of fully polarized dipoles in 2D as the ones
considered here, ${\bf p}_1$ and ${\bf p}_2$ are parallel and define a
fixed direction in space, see Fig.~\ref{petardosdeferia}. In this
case $V_d({\bf r})$ simplifies to
\begin{equation}
V_d({\bf r}) = 
{ C_{dd} \over 4\pi } \left[ { 1 - 3\lambda^2\cos^2\theta \over r^3} \right] \ ,
\label{dipdipV2D}
\end{equation}
where $\lambda=\sin\alpha$, $\alpha$ being the angle
formed by the normal to the plane and the polarization field, which is
tilted towards the $x$-axis. In this expression, $r$ and $\theta$
stand for the in-plane distance and polar angle, respectively.  Notice
that, in contrast to what happens in three dimensions, $\alpha$ is
fixed in the fully polarized system and thus $\lambda\leq 1$ is a
constant of the problem for a given $\alpha$.

One of the consequences of the form of the interaction is that it can
be either attractive or repulsive depending on the orientation with
respect to the polarization field. This fact alone triggers
interesting discussions about the static and dynamic properties of
dipolar systems. For instance, the formation of a density instability
observed in the elementary excitation spectrum as the formation of a
deep roton minimum has been widely discussed~\cite{Santos_2003,
  ODell_2003,Wilson_2008}, a feature that is absent when the
interaction is reduced to its purely isotropic limit in
2D~\cite{Mazzanti_2009} corresponding to $\lambda=0$ in
Eq.~(\ref{dipdipV2D}).  The presence of two- and/or many-body bound
states in stacks of dipolar layers has also raised interesting
questions~\cite{Klawunn_2010}.  Many other aspects including
scattering properties in
2D~\cite{Ticknor_2009,Ticknor_2010,Ticknor_2011}, pseudopotential
treatments~\cite{Kanjilal_2007} or the impact of the anisotropy on the
superfluid properties of a dipolar Bose gas~\cite{Ticknor_2010b} have
gathered interest in the recent years. Most of these aspects and many
others are covered in detail in the review article by Lahaye {\em et
  al}~\cite{Lahaye_2009}. In this work we concentrate on the case
where the interaction is always repulsive, but is still
anisotropic. That defines a critical angle $\alpha_c=0.615$ above
which $V_d({\bf r})$ starts to present negative contributions. We thus
analyze the role of the anisotropy of the interaction in situations
where the system is clearly prevented from collapse. Since $\alpha_c$
is reached at $\lambda^2=1/3$, we propose to use $\lambda^2$ as an
expansion parameter.

Despite the relevance of the anisotropic character of the interaction
in all these effects, not much effort has been put in the microscopic
description of the wave function. The anisotropy makes the different
angular momentum channels couple in a non-trivial way, and that
introduces additional degrees of freedom that must be taken into
account in a numerical simulation. At low densities, though, the
problem simplifies since the zero-energy solution of the two-body
scattering problem dominates the ground state many-body wave function
of the homogeneous gaseous phase.  In this work we provide a detailed
description of the zero-energy scattering wave function corresponding
to two dipoles moving on the plane, as a function of the polarization
angle.  Equipped with that solution we build a microscopic variational
many-body wave function that we use in a Monte Carlo simulation to
describe the static properties of a gas of polarized dipoles in 2D at
low densities.

The analysis of the low density equation of state of a gas of weakly
interacting particles has historically attracted great
interest. Corrections to the mean field prediction for
three-dimensional~\cite{Lee_1957} and
one-dimensional~\cite{Girardeau-1960} systems have been known for more
that fifty years now. The two-dimensional case has been much more
controversial as already the two-body problem presents logarithmic
divergences in the leading scattering parameters that make series
expansions difficult to carry out~\cite{Lapidus_1982,Adhikari_1986}.
In any case, the low density behavior of a gas of weakly interacting
particles in 2D has been widely discussed in the literature for the
case of isotropic interactions. One of the most remarkable properties
exhibited by these systems is the {\em universal} behavior of the
energy per particle, which admits a non--analytic series expansion in
the gas parameter $x= n a^2$, with $n$ the density and $a$ the
$s-$wave scattering length. The leading order, mean-field term in this
series has been derived by several
authors~\cite{Schick_1971,Lieb_2001} and reads
\begin{equation}
\epsilon_{mf}(x) = \left( {2ma^2 \over \hbar^2}\right) 
  {E \over N} = {4\pi x \over
  |\ln x |} \ .
\label{Elo}
\end{equation}
The detailed form of the next-to-leading correction to this expression
have been the subject of discussion and different authors proposed
different forms in the past, see for instance
Refs.~\cite{Kolomeiski_1992, Cherni_2001}.  The correct expression was
recently derived in Ref.~\cite{Mora_2009} and checked against
numerically intense Monte Carlo simulations in Ref.~\cite{Astra_2009}.
For the model system of hard disks, the mean field prediction of
Eq.~(\ref{Elo}) holds well starting at $x\sim 0.001$ and down to quite
low but still experimentally affordable values of the gas
parameter~\cite{Mazzanti_2005}.  However, no particular attention has
been paid in all these works to the special case of anisotropic
interactions.

In this article we discuss to which extent the mean field law of
Eq.~(\ref{Elo}) holds for the special case of the spatially
anisotropic dipole-dipole interaction of Eq.~(\ref{dipdipV2D}) when
the polarization angle varies between 0 and $\alpha_c$. We are
particularly interested in discerning whether the angular dependence
of the interaction has a noticeable impact on the mean-field
prediction of Eq.~(\ref{Elo}) and on other relevant ground state
properties. In order to do that, we first solve in section II the
zero-energy two-body scattering problem and obtain an expression for
the scattering length as a function of the polarization angle.  We
then use this result to build in section III a variational many-body
wave function of the Jastrow form that we use as an input to both a
variational Monte Carlo (VMC) and a diffusion Monte Carlo (DMC)
calculations from where we obtain the equation of state as a function
of $x$.  For the sake of completeness we also analyze the pair
distribution function, the static structure factor, the one-body
density matrix and the condensate fraction, and discuss how these
quantities scale on the gas parameter for different polarization
angles and densities. Finally, in section IV the main conclusions of
the work are summarized and discussed.

\pagebreak

\section{Zero energy two body problem}

In this section we develop a series expansion of the zero energy
scattering solution for two dipoles moving on the XY plane as a
function of the polarization angle $\alpha$. This is done by first
building the Green's function of the $\alpha=0$ case corresponding to
a polarization field perpendicular to the plane. This is a
particularly suitable situation since in that case the interaction is
isotropic ($\lambda=0$ in Eq.~(\ref{dipdipV2D})) and the Schr\"odinger
equation can be exactly solved. From there we carry out a series
expansion in powers of $\lambda^2$ of the solution corresponding to
the anisotropic case.  The anisotropy makes the different angular
momentum channels couple, and we report the expression for all orders
in $\lambda^2$ contributing to each partial wave. We end this section
reporting the scattering length of the dipole-dipole interaction as a
function of the polarization angle, required afterwards to analyze the
low density properties of the anisotropic many-body system.

The Hamiltonian describing the relative motion of two polarized
dipoles of mass $m$ moving on the plane reads
\begin{equation}
\hat H_2 = -{\hbar^2 \over 2 M}\nabla^2 + 
{ C_{dd} \over 4\pi } \left[ { 1 - 3\lambda^2\cos^2\theta \over r^3}
  \right]
\label{H2O}
\end{equation}
where $M=m/2$ is the reduced mass. In the following we use
dimensionless variables scaled according to the characteristic dipolar
length $r_d=m C_{dd}/4\pi\hbar^2$ and energy $\epsilon_d=\hbar^2/mr_d^2$.

The two linearly independent solutions of the zero energy
Schr\"odinger equation (SE) for the isotropic case are the building
blocks we need in order to generate the Green's function that we will
use afterwards to solve the anisotropic case. Setting $\lambda=0$, the
SE of the relative motion of the two dipoles at zero energy reduces to
\begin{equation}
  -\nabla^2\varphi+\frac{1}{r^3}\varphi = 0 \ ,
\end{equation}
and the general solution of this equation can be expanded in partial
waves as
\begin{equation}
  \varphi(r,\theta) = \sum_{n=0}^{\infty}\varphi_n(r)\cos(n\theta) \ ,
\end{equation}
where the wave function of each separate mode $\varphi_n(r)$ satisfies
\begin{equation}
  -\frac{1}{r}
  \frac{d}{dr}\left(r\frac{d\varphi_n}{dr}\right)+\left(\frac{n^2}{r^2}+
  \frac{1}{r^3}\right)\varphi_n=0 \ ,
\label{diffeqE0}
\end{equation}
which is a modified Bessel equation for $\varphi_n(2/\sqrt{r})$. The
two linearly independent solutions of Eq.~(\ref{diffeqE0}) are
$K_{2n}(2/\sqrt{r})$ and $I_{2n}(2/\sqrt{r})$~\cite{Abra_1972}, and
these two functions enter in the Green's function we write below. The
zero energy solution of the SE requires the condition
$\varphi_{2n}(r=0)$ to be zero, and that discards the
$I_{2n}(2/\sqrt{r})$ contributions. Apart from a normalization
constant, the zero energy solution of the $\lambda=0$ problem becomes
then
\begin{equation}
  \varphi_n(r) = K_{2n}\left(\frac{2}{\sqrt{r}}\right) \ .
\end{equation}
The physical solution of the isotropic case corresponds to $n=0$ as
otherwise anisotropic contributions would dominate at large distances
since $K_{2n}(2/\sqrt r)$ grows as $r^n$ when $r\to\infty$.  In 1D and
3D, the knowledge of the $E=0$ solution allows one to immediately
obtain an exact expression for the $s-$wave scattering length $a_s$.
The description of two-body scattering in 2D is more involved since
the low-energy expansion of the scattering amplitude diverges at low
energies, thus introducing additional problems not found in higher and
lower dimensions.  The scattering length $a_s$ can however be defined
to be equal to the position of the node of the asymptotic form of the
zero energy two-body wave function. This definition has the additional
advantage that can be used in any dimensions and will therefore be
adopted throughout this work~\cite{Lieb_2001,Astra_2010}.  The large
$r$ behavior of $K_0(2/\sqrt{r})$ is $-\gamma + {1\over 2}\ln(r)$ and
that yields the well known expression
\begin{equation}
  a_s=e^{2\gamma} \approx 3.17222\ldots
\label{scattlength_iso}
\end{equation}
where $\gamma$ is Euler's gamma constant.

The SE describing the anisotropic case can be cast in the form
\begin{equation}
  -\nabla^2\phi+\frac{1}{r^3}\phi=\frac{3\lambda^2\cos^2\theta}{r^3}\phi
  \ ,
\end{equation}
and the general solution for $\lambda\neq 0$ can be derived from the
Green's function corresponding to $\lambda=0$, which
fulfills the equation
\begin{equation}\label{vecGreen}
  \left(-\nabla^2+\frac{1}{r^3}\right)G(\textbf{r},\textbf{r}')=
  \delta(\textbf{r}-\textbf{r}')
  \ ,
\end{equation}
leading to
\begin{equation}
  \phi(\textbf{r}) = \varphi_0(r)+3\lambda^2\int d\textbf{y}\,
  \frac{\cos^2\theta_y}{y^3}
  G(\textbf{r},\textbf{y})\phi(\textbf{y}) 
  \label{Lippmann}
\end{equation}
with $\varphi_0(r)=K_0(2/\sqrt{r})$ the $\lambda=0$ solution as described above.

Equation~(\ref{vecGreen}) can be solved expanding the Green's function
in partial waves as before
\begin{equation}
  G(\textbf{r},\textbf{r}')=\frac{1}{2\pi}g_0(r,r')
  +\frac{1}{\pi}\sum_{n=1}^{\infty}g_{n}(r,r')\cos\left[n(\theta-\theta')\right]
\end{equation}
where
\begin{equation}
  g_n(r,r') = 
    \left\{
    \begin{array}{cc}
      2K_n\left(\frac{2}{\sqrt{r}}\right)I_n\left(\frac{2}{\sqrt{r'}}\right)
      & \qquad \text{if $r<r'$} \\  
      2I_n\left(\frac{2}{\sqrt{r}}\right)K_n\left(\frac{2}{\sqrt{r'}}\right)
      & \qquad \text{if $r>r'$} \\  
    \end{array}
    \right.
    \label{greenfuncmodes}
\end{equation}
satisfies the boundary condition $g_n(r=0,r')=g_n(r,r'=0)=0$ while
keeping it bounded at large distances. The general solution of the
Fredholm integral equation~(\ref{Lippmann}) admits a series expansion
in powers of $\lambda^2$
\begin{equation}
  \phi(\textbf{r}) = \sum_{k=0}^{\infty}\lambda^{2k}\phi^{(k)}(\textbf{r}) \ ,
\end{equation} 
where each $\phi^{(k)}(\textbf{r})$ satisfies the recurrence relation 
\begin{equation}
  \phi^{(k+1)}(\textbf{r}) = 3\lambda^2\int d\textbf{y}\,
  \frac{\cos^2\theta_y}{y^3}
  G(\textbf{r},\textbf{y})\phi^{(k)}(\textbf{y}) \ .        
\end{equation}
When $\phi^{(k)}({\bf r})$ is further expanded in partial waves
and the Bose symmetry is taken into account
\begin{equation}
  \phi^{(k)}(r,\theta) = \sum_{n=0}^{\infty}\phi^{(k)}_{2n}(r)\cos(2 n \theta) \ ,
\end{equation}
the coupling between the different angular momentum channels produced
by the $\cos^2\theta$ term of the interaction emerges and the radial
functions satisfy the following recurrence relations for even $n$
\begin{equation}
  \phi_n^{(k+1)}(r) = \frac{3\lambda^2}{4}\int_0^{\infty}dy\, \frac{g_n(r,y)}{y^2}
  \left[\phi_{n+2}^{(k)}(y)+2\phi_n^{(k)}(y)+
  \phi_{|n-2|}^{(k)}(y)\right]
  \label{recurrence_phink}
\end{equation}
that can be solved iteratively starting from
$\phi_0^{(0)}(r)=\varphi_0(r)$.  From this expression one sees that by
adding successive orders in $\lambda^2$ to the series expansion of
$\phi({\bf r})$, more angular momentum channels couple together.  As
in the regime considered the interaction is fully repulsive,
$\lambda~<~1/\sqrt{3}$ and that makes $\lambda^2$ a small parameter
that we can use in a series expansion of the solution.  In fact, it
can be shown from the previous expressions that $\phi_{2n}^{(k)}(r)=0$
for $2n>k$, and that therefore the lowest order contribution to the
$n$-th mode is $\lambda^{2n}$.  By adding $\phi_{2n}^{(k)}(r)$ for all
$k$ and fixed $n$ one recovers $\phi_{2n}(r)$, the complete $2n-th$
mode contribution to $\phi({\bf r})$.  We thus find
\[
\phi_{2n}(r) = \sum_{k=n}^\infty \lambda^{2k} \phi_{2n}^{(k)}(r) \ ,
\]
which means that, up to a given order $\lambda^{2k}$, the total wave
function $\phi({\bf r})$ has contributions coming only from channels
$n=0,2,\ldots, 2k$.

\begin{figure}[!t]
  \centering
  \includegraphics[width=0.5\textwidth]{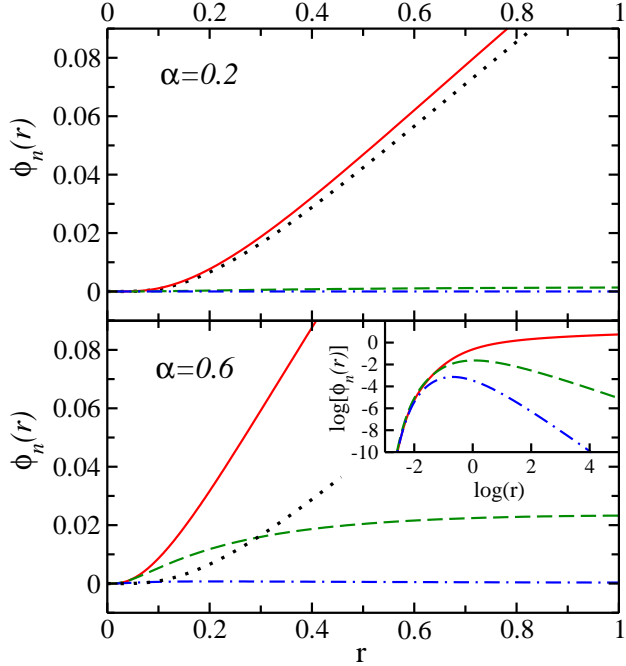}
  \caption{(Color online) Lowest order partial wave contributions to
    the zero energy two-body scattering wave function. The upper and
    lower panels show $\phi_0^{(0)}=K_0(2/\sqrt{r})$ (black dotted
    line), 
    $\phi_0(r)$ (red solid line), 
    $\phi_2(r)$ (green dashed line) and $\phi_4(r)$ 
    (blue dot-dashed line) for the polarization angles $\alpha=0.2$ and
    $\alpha=0.6$.  
    The inset show the $\alpha=0.6$ curves in logarithmic scale.}
  \label{fig_modes}
\end{figure}

Figure~\ref{fig_modes} shows the lowest order partial waves
contributions corresponding to the two polarization angles
$\alpha=0.2$ (upper panel) and $\alpha=0.6$ (lower panel), the latter
being the largest angle considered in this work. The latter angle is
fairly close to the critical angle $\alpha_c=0.615$ where the
interaction ceases to be purely repulsive.  At $\alpha=0.6$ the
contribution of the $n\neq 0$ modes is expected to be larger than for
any lower angle. This is clearly seen from the figure, where the
$\lambda^2$ corrections to the $n=0$ and $n=2$ partial waves are shown
(red solid and green dashed lines), as well as the leading $\lambda^4$
correction corresponding to the $n=0$ mode. It is clear from
Eqs.~(\ref{greenfuncmodes}) to~(\ref{recurrence_phink}) and the
positiveness of the modified Bessel functions that every radial
contribution $\phi_{2n}^{(k)}(r)$ to the two-body wave function is
also positive, as seen for the lowest mode contributions in the
figure.  It is also apparent that the lower the angle, the smaller the
correction to the $\alpha=0$ solution $\varphi(r)$ is, as
expected. Despite the fact that the series expansion of the two-body
solution $\phi({\bf r})$ is in general alternating due to the cosine
terms, the total two-body wave function does not change sign as the
interaction is everywhere repulsive, thus making the $E=0$ scattering
solution be the ground state.

\begin{figure}
  \centering
  \includegraphics[width=0.5\textwidth]{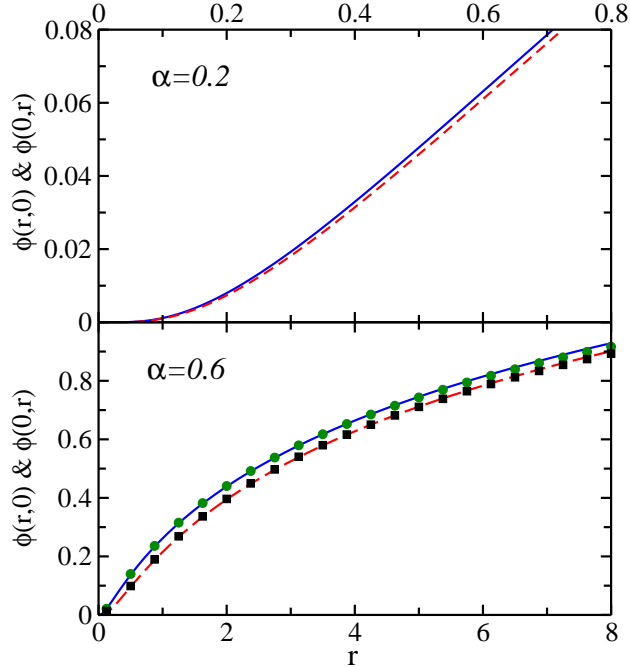}
  \caption{(Color online) Cuts of the zero energy two-body scattering
    wave function describing the relative motion of two dipoles. The
    blue solid line and the red dashed line correspond to the cuts
    along the $x$ and $y$ axes, respectively. The green circles and
    black squares are the prediction of the optimal Jastrow factor
    obtained from the solution of the HNC/0 Euler equations for a
    value of the gas parameter $x=0.01$.}
    \label{fig_2body_xy}
\end{figure}

The effect of the anisotropy on the ground state wave function is seen
in Fig.~\ref{fig_2body_xy} where two cuts, one along the $x$-axis
(contained in the plane formed by the polarization vector and the
$z$-axis), and another in the perpendicular direction ($y$-axis) are
depicted for the two polarization angles $\alpha=0.2$ and
$\alpha=0.6$.  These cuts coincide with the directions where the
interaction is least and most repulsive, respectively. As it can be
seen, anisotropic effects are visible in both cases but are more
pronounced at high polarization angles .  The inset in
Fig.~\ref{fig_modes} shows that the $m=0$ mode dominates at very large
distances as expected, making the asymptotic wave function be
isotropic.  From $\phi_0(r)$ one can extract the scattering length
$a_s(\lambda)$ of the anisotropic dipolar interaction, which is given
by the node of its asymptotic $r\to\infty$ form.  An analytic
approximation to $a_s(\lambda)$ can be easily obtained recalling that
every mode $\phi_n(r)$ contributes to order $\lambda^{2n}$ and that
therefore the anisotropy enters at order $\lambda^2$.  Direct
inspection of the modes expansion of the the Schr\"odinger equation
for $\phi({\bf r})$ reveals that $\phi_0(r)$ and $\phi_2(r)$ are
related according to
\begin{equation}
  -\frac{1}{r}\frac{d}{dr}\left(r\frac{d\phi_0}{dr}\right)+
  \frac{1}{r^3}\phi_0 = \frac{3\lambda^2}{2r^3}
  \left[\phi_0(r) + {1\over 2}\,\phi_2(r)\right] \ .
\end{equation} 
An approximation of order $\lambda^2$ to $a_s(\lambda)$ can be
obtained by keeping only the $\phi_0(r)$ mode on the right hand side
of this equation
\begin{equation}
  -\frac{1}{r}\frac{d}{dr}\left(r\frac{d\phi_0^{(0)}}{dr}\right)+
  \frac{1}{r^3}\phi_0^{(0)} = \frac{3\lambda^2}{2r^3}\phi_0^{(0)}(r) \ ,
\end{equation} 
which once again is a modified Bessel equation with the 
general solution
\begin{equation}
  \phi_0^{(0)}(r) = \mathcal{N} \left[
  K_0\left(2\sqrt{\frac{1-\frac{3\lambda^2}{2}}{r}}\right) + 
  B(\lambda) I_0\left(2\sqrt{\frac{1-\frac{3\lambda^2}{2}}{r}}\right)
  \right]
\end{equation}
with $\mathcal{N}$ a normalization constant. In this expression
$B(\lambda)$ is an unknown function of $\lambda^2$ according to the
parity of the Hamiltonian under the $\lambda\to-\lambda$
transformation. Furthermore, $B(0)=0$ so that one recovers the
isotropic solution given in Eq.~(\ref{scattlength_iso}).  Hence
$B(\lambda)=b_2 \lambda^ 2$ to order $\lambda^2$, with $b_2$ a
constant.  In the asymptotic $r\to\infty$ regime, $I_0\approx 1$ and
one can compare the expansion to order $\lambda^2$ of the above
expression to the expression of $\phi_0(r)$ to the same order obtained
from the integration of the Green's function done before.  This yields
$b_2=0$ and one has
\begin{equation}
  \phi_0(r\to\infty) \to
  \left. 
  K_0\left(2 \sqrt{\frac{1-\frac{3\lambda^2}{2}}{r}}\right)
  \right|_{r\to\infty}
  \approx
  \frac{1}{2}\ln\frac{r}{a_s(\lambda)}
\end{equation}
with $a_s(\lambda)$ the s-wave scattering length 
\begin{equation}
  a_s(\lambda) = e^{2\gamma}\left(1-\frac{3\lambda^2}{2}\right) \ .
  \label{scatt_length_approx}
\end{equation}

\begin{figure}
  \centering
  \includegraphics[width=0.6\textwidth]{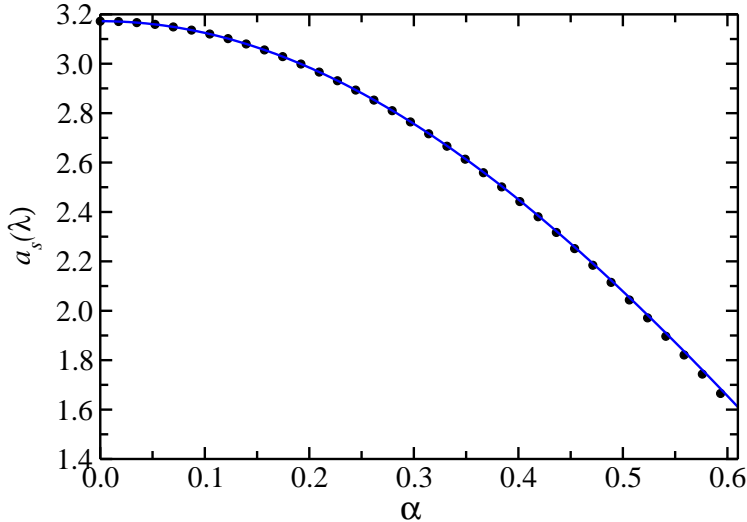}
  \caption{(Color online) $s$-wave scattering length 
    $a_s(\lambda=\sin(\alpha))$ as a 
    function of the polarization angle $\alpha$. 
    The blue solid line and the black dots correspond
    to the exact numerical solution and the $\lambda^2$ approximation of 
    Eq.~(\ref{scatt_length_approx}), respectively.}
    \label{scat_length}
\end{figure}

This expression is accurate up to order $\lambda^2$, so one could
expect it to provide a reliable prediction only at small polarization
angles.  This turns out not to be the case, and in
Fig.~\ref{scat_length} we show the comparison of this approximation to
the exact result obtained by numerically finding the node of the
asymptotic $m=0$ wave function, which is isotropic and dominates the
large distance behavior of $\phi({\bf r})$.  As can be seen from the
figure, the approximation works surprisingly well up to the critical
angle $\alpha_c$ where the interaction ceases to be fully repulsive.
Deviations increase with increasing polarization angle, but even at
$\alpha=\alpha_c$ the separation between the approximation in
Eq.~(\ref{scatt_length_approx}) and the exact numerical estimation is
less than a $3\%$.

\section{Many-Body description}

In this section we extend the previous discussion and analyze, using
DMC and VMC methods, the most relevant ground state properties of an
homogeneous and anisotropic gas of bosonic dipoles with a polarization
angle $\alpha$ lower than the critical angle $\alpha_c=0.615$. We
stick to the low density limit were the system, characterized by a
fully repulsive and anisotropic interaction, remains in a stable,
gaseous phase.  One of the most relevant quantities to analyze at low
densities is the total energy per particle of the gas and its
universal scaling properties.  Quite a lot of work has been devoted in
the past to that question, including both
3D~\cite{Lee_1957,Fetter_1971,Giorgini_1999, Mazzanti_2003} and
2D~\cite{Schick_1971,Lieb_2001,Kolomeiski_1992,Cherni_2001,
  Mora_2009,Astra_2009,Mazzanti_2005} systems.  However, little has
been discussed about the same properties in anisotropic systems as the
dipolar gas considered here.  We analyze the impact of the
polarization angle $\alpha$ formed by the dipoles on the universality
scaling law exhibited by other isotropic, short ranged interactions.

The Hamiltonian of the system of fully polarized dipoles,
written in the same dipolar units used in the previous section, becomes
\begin{equation}
H = -{1\over 2}\sum_{j=1}^N \nabla_j^2 + 
\sum_{i<j} {1-3\lambda^2 \cos^2 \theta_{ij} \over r_{ij}^3} \ ,
\label{hamiltonian}
\end{equation}
with $\lambda=\sin\alpha$, and $r_{ij}$ and $\theta_{ij}$ the distance
and angle formed by dipoles $i$ and $j$, respectively, measured on the
plane.

The leading ground state quantities describing the low density static
properties of the system can be obtained using different techniques.
In this work we stick to diffusion and variational Monte Carlo
methods, widely used nowadays in the analysis of weakly and strongly
correlated systems.  Variational Monte Carlo samples stochastically a
trial wave function and yields an upper bound to the real ground
state energy of the system. On the other hand, diffusion Monte Carlo
uses also a trial wave function to guide the sampling process but
removes the contributions from excited states to finally yield
statistically exact energies. In both cases, however, a suitable
variational wave function is required.  The quality of the results is
directly related to the quality of the wave function employed in the
VMC case, while DMC is far less demanding and any reasonable guiding
function can be used as long as it is not orthogonal to the true
ground state. But even in DMC a high quality wave function makes the
method converge faster and with smaller variance towards the exact
result.  Consequently, seeking for a good trial many-body wave
function $\Psi({\bf r}_1, {\bf r}_2, \ldots, {\bf r}_N)$ is always
desirable.

In this work we use a model wave function of the Jastrow form
\begin{equation}
\Psi({\bf r}_1, {\bf r}_2, \ldots, {\bf r}_N) = 
\prod_{i<j} f({\bf r}_{ij}) \ ,
\label{jastrow-b1}
\end{equation}
where the two-body correlation factor 
$f({\bf r}_{ij})=f({\bf r}_i-{\bf r}_j)$ depends on the position vector 
linking particles $i$ and $j$. One significant difference between this
Jastrow factor and the ones most commonly employed in the analysis of
other condensed matter systems is that, due to the anisotropic
character of the interaction, $f({\bf r})$ depends explicitly on the whole
${\bf r}$ vector and not only on its magnitude.  In this way, the wave
function in Eq.~(\ref{jastrow-b1}) describes an homogeneous but
anisotropic system as the one under study.

At low densities, the zero-energy scattering solution of the two-body
problem greatly influences the structural properties of the gas.  For
that reason we use as a Jastrow factor the anisotropic solution of the
relative motion of two dipoles on the plane derived in the previous
section. The $n=0$ mode of this wave function is matched at some
healing distance $\xi$ with the symmetrized form of a phononic wave
function $f_\xi(r)=\exp(-C/r)$~\cite{Reatto_1967}, taking both $\xi$
and $C$ as variational parameters and imposing the continuity of
$f(r)$ and $f'(r)$ at $r=\xi$.  The $n>0$ modes of the two-body
problem, inducing the anisotropy of $\Psi({\bf r}_1, {\bf r}_2,
\ldots, {\bf r}_N)$ in Eq.~(\ref{jastrow-b1}), decay to zero at large
distances and so their influence at the boundaries of the simulation
box is marginally small.  Alternatively, the optimal Jastrow factor
corresponding to the many-body problem can be obtained from the
solution of the HNC/0 Euler-Lagrange equations~\cite{Kroky_1998}.
Although not exact, the optimized HNC/0 solution gives an accurate
variational description of quantum Bose systems and captures most of
the short and long range features of the exact ground state wave
function.  For the sake of comparison, we also show in
Fig.~\ref{fig_2body_xy} the optimized HNC/0 Jastrow factor (black and
green symbols) at $x=0.01$ and polarization angle $\alpha=0.6$. The
comparison indicates that the two-body solution provides an accurate
description of the two-body correlation factor, which becomes even
better as the gas parameter is reduced.  We have checked that the
HNC/0 Jastrow factor and the solution of the two-body problem are in
very good agreement in the whole range of gas parameter values
considered in this work.

\begin{table}
\begin{center}
\hspace*{-12mm}
{\tabcolsep=1.5mm
\begin{tabular}{|c|c|c|c|c|c|c|} \cline{2-7}
\multicolumn{1}{c|}{} & \multicolumn{2}{|c|}{$\alpha=0.2$} & 
\multicolumn{2}{|c|}{$\alpha=0.4$} & \multicolumn{2}{|c|}{$\alpha=0.6$} \\ \hline
$x$ & DMC & VMC & DMC & VMC& DMC & VMC \\ \hline
$10^{-7}$ & $4.271(61)\!\cdot\!10^{-9}$ & $4.268(92)\!\cdot\!10^{-9}$ & 
$6.469(62)\!\cdot\!10^{-9}$ & $6.490(24)\!\cdot\!10^{-9}$ & 
$1.414(62)\!\cdot\!10^{-8}$ & $1.429(75)\!\cdot\!10^{-8}$ \\
$5\!\cdot\!10^{-7}$ & $2.386(24)\!\cdot\!10^{-8}$ & $2.389(90)\!\cdot\!10^{-8}$ &
$3.602(70)\!\cdot\!10^{-8}$ & $3.633(91)\!\cdot\!10^{-8}$ &
$7.888(15)\!\cdot\!10^{-8}$ & $7.931(45)\!\cdot\!10^{-8}$ \\
$10^{-6}$ & $5.030(32)\!\cdot\!10^{-8}$ & $5.044(91)\!\cdot\!10^{-8}$ &
$7.614(21)\!\cdot\!10^{-8}$ & $7.631(36)\!\cdot\!10^{-8}$ &
$1.664(50)\!\cdot\!10^{-7}$ & $1.690(86)\!\cdot\!10^{-7}$ \\
$5\!\cdot\!10^{-6}$ & $2.868(24)\!\cdot\!10^{-7}$ & $2.874(23)\!\cdot\!10^{-7}$ & 
$4.317(70)\!\cdot\!10^{-7}$ & $4.360(89)\!\cdot\!10^{-7}$ &
$9.448(93)\!\cdot\!10^{-7}$ & $9.472(85)\!\cdot\!10^{-7}$ \\
$10^{-5}$ & $6.105(64)\!\cdot\!10^{-7}$ & $6.135(87)\!\cdot\!10^{-7}$ &
$9.271(41)\!\cdot\!10^{-7}$ & $9.312(22)\!\cdot\!10^{-7}$ &
$2.032(90)\!\cdot\!10^{-6}$ & $2.011(92)\!\cdot\!10^{-6}$ \\
$5\!\cdot\!10^{-5}$ & $3.584(31)\!\cdot\!10^{-6}$ & $3.596(27)\!\cdot\!10^{-6}$ & 
$5.405(15)\!\cdot\!10^{-6}$ & $5.450(94)\!\cdot\!10^{-6}$ &
$1.180(40)\!\cdot\!10^{-5}$ & $1.199(81)\!\cdot\!10^{-5}$ \\
$10^{-4}$ & $7.744(61)\!\cdot\!10^{-6}$ & $7.768(72)\!\cdot\!10^{-6}$ & 
$1.170(41)\!\cdot\!10^{-5}$ & $1.177(30)\!\cdot\!10^{-5}$ &
$2.542(88)\!\cdot\!10^{-5}$ & $2.579(84)\!\cdot\!10^{-5}$ \\
$5\!\cdot\!10^{-4}$ & $4.734(49)\!\cdot\!10^{-5}$ & $4.757(48)\!\cdot\!10^{-5}$ & 
$7.124(93)\!\cdot\!10^{-5}$ & $7.205(59)\!\cdot\!10^{-5}$ &
$1.555(62)\!\cdot\!10^{-4}$ & $1.567(73)\!\cdot\!10^{-4}$ \\
$10^{-3}$ & $1.046(16)\!\cdot\!10^{-4}$ & $1.051(31)\!\cdot\!10^{-4}$ & 
$1.577(33)\!\cdot\!10^{-4}$ & $1.590(58)\!\cdot\!10^{-4}$ &
$3.425(30)\!\cdot\!10^{-4}$ & $3.467(23)\!\cdot\!10^{-4}$ \\
$5\!\cdot\!10^{-3}$ & $6.776(61)\!\cdot\!10^{-4}$ & $6.807(74)\!\cdot\!10^{-4}$ &
$1.018(90)\!\cdot\!10^{-3}$ & $1.029(58)\!\cdot\!10^{-3}$ &
$2.222(51)\!\cdot\!10^{-3}$ & $2.240(26)\!\cdot\!10^{-3}$ \\
$10^{-2}$ & $1.532(20)\!\cdot\!10^{-3}$ & $1.551(31)\!\cdot\!10^{-3}$ & 
$2.316(31)\!\cdot\!10^{-3}$ & $2.337(23)\!\cdot\!10^{-3}$ &
$5.036(55)\!\cdot\!10^{-3}$ & $5.067(97)\!\cdot\!10^{-3}$ \\
$5\!\cdot\!10^{-2}$ & $1.077(11)\!\cdot\!10^{-2}$ & $1.085(29)\!\cdot\!10^{-2}$ & 
$1.616(9)\!\cdot\!10^{-2}$ & $1.634(18)\!\cdot\!10^{-2}$ &
$3.517(74)\!\cdot\!10^{-2}$ & $3.544(62)\!\cdot\!10^{-2}$ \\
$10^{-1}$ & $2.534(29)\!\cdot\!10^{-2}$ & $2.572(67)\!\cdot\!10^{-2}$ &
$3.774(42)\!\cdot\!10^{-1}$ & $3.840(66)\!\cdot\!10^{-2}$ &
$8.235(21)\!\cdot\!10^{-2}$ & $8.292(21)\!\cdot\!10^{-2}$ \\
$5\!\cdot\!10^{-1}$ & $1.947(14)\!\cdot\!10^{-1}$ & $1.962(54)\!\cdot\!10^{-1}$ & 
$2.908(28)\!\cdot\!10^{-1}$ & $2.938(41)\!\cdot\!10^{-1}$ &
$6.311(33)\!\cdot\!10^{-1}$ & $6.347(32)\!\cdot\!10^{-1}$ \\ \hline
\end{tabular}
\caption{DMC and VMC energies per particle as a function of the gas parameter $x=n a^2$.}
\label{tab-E}
}
\end{center}
\end{table}

\begin{figure}
  \centering
  \includegraphics[width=0.6\textwidth]{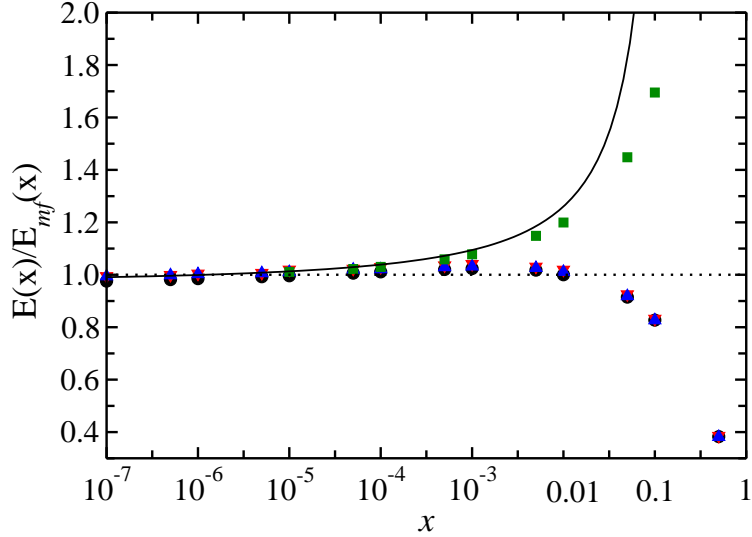}
  \caption{(Color online) Ratio of the energy per 
    particle of the gas of dipoles for
    different polarization angles to the mean field
    prediction of Eq.~(\ref{Elo}).  Black circles, red triangles and blue
    triangles correspond to $\alpha=0.2, 0.4$ and $0.6$,
    respectively. The green squares are the optimized HNC/EL
    energies for hard disks of Ref.~\cite{Mazzanti_2003}, while the
    solid line is the universal curve of Ref.~\cite{Astra_2010}. The
    dotted line corresponds to the mean field prediction.}
  \label{fig_E}
\end{figure}

Table~\ref{tab-E} lists both the VMC and DMC energies obtained from
the Jastrow trial wave function of Eq.~(\ref{jastrow-b1}) for the
polarization angles $\alpha=0.2, 0.4$ and $0.6$.  Notice that the
energies in the table are given for fixed $x$ and different
polarization angles, and since the scattering length varies with
$\alpha$, the densities change accordingly.  A direct measure of the
quality of the variational model is given by the separation between
these two measures (VMC and DMC), and one can check that the relative
difference in energies is always of the order of $1\%$ or $2\%$. Other
than that, the energy is an increasing function of the gas parameter
that yields appreciably different results for different polarization
angles.  These energies can be used to check the influence of the
anisotropic character of the dipolar interaction on the universality
scaling property fulfilled by the energy per particle of homogeneous
and isotropic systems in 2D. In order to do that, one has to express
the total energy per particle in units of $\hbar^2/2m a^2$ with $a$
the scattering length. This is achieved multiplying the energies in
Table~\ref{tab-E} (expressed in dipolar units) by $2 a_s^2(\lambda)$,
with $a_s(\lambda)$ the scattering length for the corresponding
polarization angle. Figure~\ref{fig_E} shows the ratio of the energy
per particle in units of $\hbar^2/2m a^2$ to the mean field prediction
of Eq.~(\ref{Elo}) for the three polarization angles $\alpha=0.0, 0.4$
and $0.6$. As it can be seen, expressed in scattering length units,
all curves corresponding to different polarization angles merge into a
single curve, with very small deviations that are not easily resolved
even at the highest values of gas parameters $x$ considered in this
work.  That means that the anisotropy of the interaction, present in
the wave function, does not appreciably affect the energy per particle
in the low density regime analyzed in this work.  We conclude that the
difference in energy values shown in Table~\ref{tab-E} for fixed $x$
and varying polarization angles are to be mostly attributed to the
different density $n=x/a^2$ in each case.

Figure~\ref{fig_E} also shows the universal curve including beyond
mean field effects of Ref.~\cite{Astra_2009} and the optimized HNC/0
prediction for a gas of hard disks of Ref.~\cite{Mazzanti_2005}.  As
it can be seen, the universal and the hard disks curves are close
to each other while the dipole curves remain closer to the mean field
prediction $\epsilon_{mf}(x)$ as the gas parameter is raised. Starting
at $x\sim 0.05$ the dipole curves bend downwards and the energy
deviates significantly from $\epsilon_{mf}(x)$.  In any case, it is
clear from the figure that the universality regime where the energy
per particle depends only on the gas parameter of the interaction is
left much before anisotropic effects have an appreciable impact on the
energy of the dipolar gas.

\begin{figure}
  \centering
  \includegraphics[width=0.7\textwidth]{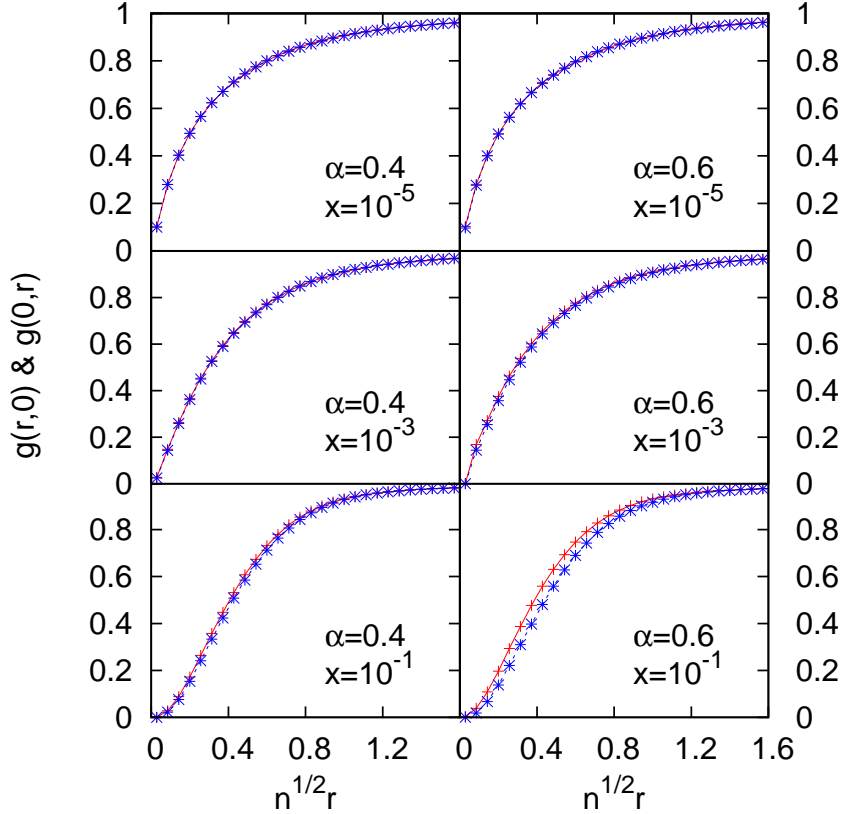}
  \caption{(Color online) Pair distribution function for 
  $\alpha=0.4$ and $\alpha=0.6$ and three values of the 
  gas parameter. The red and blue curves show the two cuts 
  $g(r,0)$ and $g(0,r)$, respectively.}
  \label{fig_gxy}
\end{figure}

\begin{figure}
  \centering
  \includegraphics[width=0.8\textwidth]{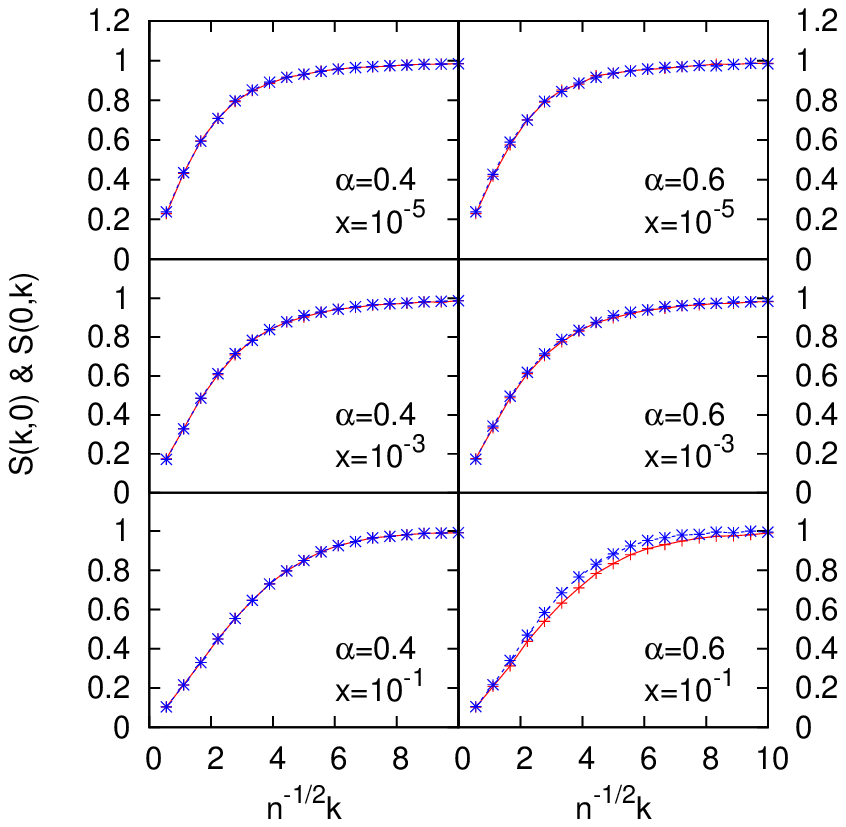}
  \caption{(Color online) Static structure function    for polarization
  angles $\alpha=0.4$ and $\alpha=0.6$ for $x=10^{-5}, 10^{-3}$ and 
  $10^{-1}$. The red and blue curves show the two cuts 
  $S(k,0)$ and $S(0,k)$, respectively.}
  \label{fig_skxky}
\end{figure}

The anisotropic character of the dipolar interaction has a direct
influence on the ground state wave function that is reflected in the
ground state expectation value of any many-body operator.
Figures~\ref{fig_gxy} and~\ref{fig_skxky} show pure DMC
estimations~\cite{Casu_1995} of the pair distribution function $g({\bf
  r})$ and its Fourier transform, the static structure factor $S({\bf
  k})$, for two values of the polarization angle $\alpha=0.4$ and
$\alpha=0.6$ (left and right panels), and three values of the gas
parameter $x=10^{-5}, 10^{-3}$ and $10^{-1}$ (top to bottom).  Notice
that in both figures the horizontal axis has been scaled with the
square root of the density for a better comparison.  Due to the
symmetries of the Hamiltonian, the complete $g(\bf r)$ and $S({\bf
  k})$ functions vary continuously on the plane but the pattern on the
first quadrant is repeated and reflected on the other three. The
figures show only the two cuts along the perpendicular and parallel
directions with respect to the polarization plane, corresponding to
the lines where the interaction is most and least repulsive,
respectively. As it can be seen, and in agreement with what one would
expect, the effect of the anisotropy is more clearly seen at higher
polarization angles and for large values of the gas parameter, being
maximal for $\alpha=0.6$ and $x=10^{-1}$.  For fixed $\alpha$ the
separation between $g(r,0)$ and $g(0,r)$ is enhanced with increasing
$x$, as happens with $S(k,0)$ and $S(0,k)$.  Accordingly and for a
given $x$, the separation between the curves also increases when the
polarization angle is raised.  In any case it is remarkable how the
anisotropy present in $g({\bf r})$ and $S({\bf k})$ changes with the
polarization angle as can be seen from the figures at large $x$, while
the total energies per particle are almost the same when properly
scaled with the scattering length. This points towards a delicate
balance between the kinetic and potential contributions, which change
with $\alpha$ but keep their sum constant once expressed in scattering
length units.

The last quantity analyzed in this work is the one-body density matrix
$\rho_1({\bf r}_1, {\bf r}_1')$, which provides a measure of the
overlap between two instances of the ground state wave function when
one particle is shifted from its initial position at ${\bf r}_1$ to a
new position at ${\bf r}_1'$
\begin{equation}
\rho_1({\bf r}_1, {\bf r}_1') = 
N {
\int d{\bf r}_2 \cdots {\bf r}_N
\Psi_0({\bf r}_1, {\bf r}_2, \ldots, {\bf r}_N)
\Psi_0({\bf r}_1', {\bf r}_2, \ldots, {\bf r}_N)
\over
\int d{\bf r}_1 d{\bf r}_2 \cdots {\bf r}_N
\Psi^ 2_0({\bf r}_1, {\bf r}_2, \ldots, {\bf r}_N)
} \ .
\end{equation}
In the case of translationally invariant systems as the one under
study, the one-body density matrix depends on its arguments only
through their difference and thus $\rho_1({\bf r}_1, {\bf r}_1') =
\rho_1({\bf r}_1-{\bf r}_1',0)\equiv \rho_1({\bf r}_{11'})$
Additionally, if the interaction is isotropic, $\rho_1$ depends only
on the magnitude of its argument $r_{11'}=|{\bf r}_{11'}|$ and its
large-$r_{11'}$ limit measures directly the condensate fraction $n_0$
which is proportional to the number of particles in the Bose-Einstein
condensate.  In the present case, however, the system is homogeneous
but not isotropic so $\rho_1({\bf r}_{11'})$ will depend on the
direction of ${\bf r}_{11'}$.  Due to translational invariance,
though, momentum is still a good quantum number and one expects
condensation to appear at the zero momentum state. In that sense one
can still write the relation between $\rho_1({\bf r}_{11'})$ and the
momentum distribution in the form
\begin{equation}
\rho_1({\bf r}_{11'}) = \rho n_0 + {1\over (2\pi)^2} \int d{\bf k}\,
e^{i{\bf k}\!\cdot\!{\bf r}_{11'}} \tilde n({\bf k})
\label{rho1nk-b1}
\end{equation}
where $\tilde n({\bf k})$ is the momentum distribution of the non-condensate atoms.
The one-body density matrix of the anisotropic gas of Bose dipoles can be further
expanded in partial waves
\begin{equation}
\rho_1({\bf r}) = \sum_{m=0}^\infty 
\rho_{1m}(r) \cos(2m\,\theta) \ ,
\label{rho_modes}
\end{equation}
with $\rho_{1m}(r)$ the radial function corresponding to the $m$-th 
mode contribution. Notice that, as before, the Bose symmetry restricts the 
previous sum to even modes only.

\begin{figure}
  \centering
  \includegraphics[width=0.8\textwidth]{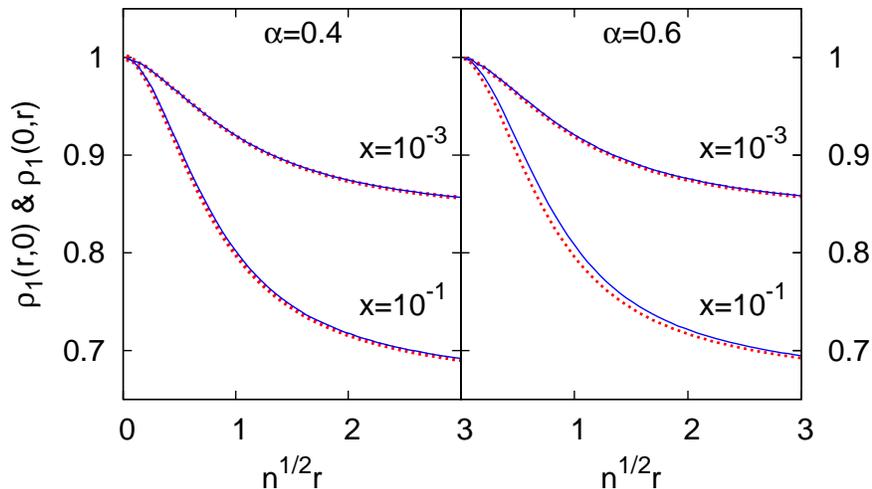}
  \caption{(Color online) Cuts of the one-body density matrix 
  along the $x$ (red dotted lines) and $y$ (blue lines) axes, for the
  gas parameter values $x=10^{-3}$ and $x=10^{-1}$ (top to bottom).
  The curves on the left and right panels correspond to $\alpha=0.4$ and
  $\alpha=0.6$, respectively.}
  \label{fig_rho1}
\end{figure}

Once enough modes $\rho_{1m}(r)$ are known, one can reconstruct the
complete one-body density matrix for all points in the plane.  In
particular, the cuts along the two directions parallel and
perpendicular to the polarization plane, corresponding to $\theta=0$
and $\pi/2$ in Eq.~(\ref{rho_modes}), turn out to be particularly easy
to evaluate
\begin{equation}
\rho_1(r,0) = \sum_{m=0}^\infty \rho_{1m}(r)
\,\,\,\,\, , \,\,\,\,\,
\rho_1(0,r) = \sum_{m=0}^\infty (-1)^m\rho_{1m}(r)
\label{rho1_cuts}
\end{equation}
and display the maximum difference two cuts along different directions
can take at the low densities considered in this work.
Figure~\ref{fig_rho1} shows the parallel and perpendicular cuts of
$\rho_1({\bf r})$ for the polarization angles $\alpha=0.4$ and
$\alpha=0.6$ (left and right panels).  The upper and lower curves
correspond to the gas parameter values $x=10^{-3}$ and $x=10^{-1}$,
respectively. As before, the coordinates on the horizontal axis have
been scaled with the density.  Similarly to what happens to the other
quantities analyzed, only at the highest gas parameter values the
effects of the anisotropy start to be visible.  This stresses once
again the minor role played by the anisotropy at low densities, even
in a non-diagonal quantity like $\rho_1({\bf r}_{11'})$.

The most significant differences in the one-body density matrix for
different values of the gas parameter appear at large distances, where
$\rho_1({\bf r}_{11'})$ reaches an asymptotic value that can be
identified with the condensate fraction $n_0$ in isotropic systems.
When the anisotropic character of the interaction is taken into
account, the presence of higher order partial waves in
Eqs.~(\ref{rho_modes}) and~(\ref{rho1_cuts}) could in principle change
this behavior, making the limiting value of $\rho_1({\bf r}_{11'})$
depend on the direction.  The role of the different partial waves in
that limit can be determined by looking at the momentum distribution
of the system, which can be obtained from $\rho_1({\bf r}_{11'})$ by
looking at the inverse of Eq.~(\ref{rho1nk-b1})
\begin{equation}
\tilde n({\bf k}) = \int_0^\infty dr\,r\int_0^{2\pi} d\theta\,
e^{i k r\cos(\theta-\varphi)} \left[
\Big( \rho_{10}(z)- \rho n_0\Big) + 
\sum_{m=1}^\infty \rho_{1m}(r) \cos(2m\theta) \right] \ ,
\end{equation}
with $\varphi$ the angle formed by ${\bf k}$ and the $x$-axis. 
Changing variables $\alpha=\theta-\varphi$, using the Jacobi-Anger
expansion of a plane wave in Bessel functions
\begin{equation}
e^{i k z \cos\alpha} = 
J_0(k z) + 2\sum_{m=1}^\infty i^m J_m(k z) \cos(m\alpha) 
\end{equation}
and taking into account the orthogonality of the 
cosine functions in the range $[0,2\pi]$, one finally finds
\begin{equation}
\tilde n({\bf k}) = 
2\pi\int_0^\infty J_0(kr) \Big(\rho_{10}(r)-n_0\Big) r dr + 
2\pi\sum_{m=1}^\infty (-1)^m \cos(2m\varphi)
\int_0^\infty J_{2m}(kr) \rho_{1m}(r) 
\,r dr 
\label{nkmodes-b1}
\end{equation}
where the first term on the right is isotropic and constitutes 
the $m=0$ mode of $\tilde n({\bf k})$, while the other terms
stand for the $m>0$ contributions. Notice once again that only
even modes appear in this expansion.

Requiring $\tilde n({\bf k})$ to be finite for all values of 
${\bf k}$ implies all integrals appearing in Eq.~(\ref{nkmodes-b1}) 
to be finite, a constraint that can only be fulfilled when the functions 
multiplying the Bessel functions decay to zero at large distances.
This condition particularly means that $n_0$ can be obtained as
the large $r$ limit of the $m=0$ mode of the one-body density matrix, 
which is the isotropic contribution to $\rho_1({\bf r}_{11'})$. 
This is the direct generalization of the usual procedure employed to 
determine $n_0$ in homogeneous and isotropic systems.

\begin{figure}
  \centering
  \includegraphics[width=0.8\textwidth]{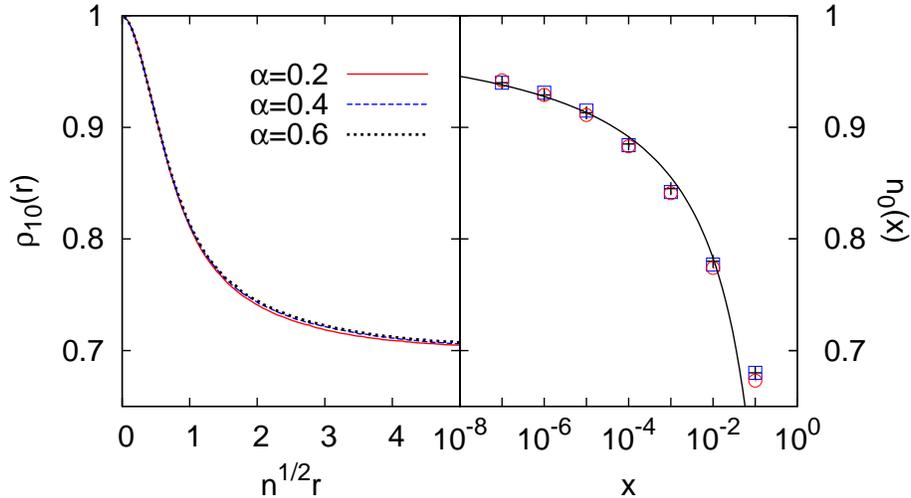}
  \caption{(Color online) Left panel: isotropic ($m=0$) contribution to the
  one-body density matrix at $x=0.1$ for the three polarization angles 
  $\alpha=0.2, 0.4$ and $0.6$ (red solid, blue dashed and black dotted lines, respectively).
  Right panel: Condensate fraction $n_0$ as a function of the gas parameter
  for $\alpha=0.2, 0.4$ and $0.6$, compared with 
  the Bogoliubov prediction (black line). The color coding for the symbols is the 
  same as in the left panel.}
  \label{fig_n0}
\end{figure}

Figure~\ref{fig_n0} shows on the left panel the $m=0$ mode
contribution $\rho_{10}(r)$ for the three polarization angles
$\alpha=0.2, 0.4$ and $0.6$ in terms of the scaled distances $n^{1/2}
r$ for $x=0.1$. As can be seen from the figure, all three curves are
hardly distinguishable, stressing once again that to a large extent
the physics is governed by the scattering length, which makes the
density change for different polarization angles when $x$ is
fixed. The right panel in the figure shows the condensate fraction as
a function of the gas parameter $x=n a^2$, obtained from the
$r\to\infty$ limit of a fit to the long range asymptotic limit of the
$m=0$ partial wave contribution of the one-body density matrix. Up to
the highest value of $x$ considered all three cases yield nearly the
same prediction within statistical errors, while differences start to
be significant only at $x\approx 0.1$.  Therefore, the scaling on the
gas parameter is preserved although moving from $\alpha=0.2$ to
$\alpha=0.6$ for fixed $x$ implies a change in density by almost a
factor of 2. The figure also shows the Bogoliubov prediction for an
isotropic gas of weakly interacting 2D bosons
\begin{equation}
n_0(x) = 1 - {1\over |\ln x|}
\label{Bogo_n0}
\end{equation}
which agrees reasonably well with the Monte Carlo prediction up to
$x\approx 0.01$ where particle correlations seem to deplete the
condensate less effectively than the mean field model.

\section{Summary and Conclusions}

To summarize, in this work we have described the ground state
properties of a gas of fully polarized Bose dipoles moving on the XY
plane, where the polarization field forms an angle $\alpha$ with the
normal direction. The projection of the polarization vector on the XY
plane defines the $x$-axis, where the potential is softer than in any
other direction. In this context, the dipole-dipole interaction
defines a critical angle $\alpha_c\approx 0.615$ where the potential
starts to have attractive contributions.  We have solved the zero
energy two-body scattering problem by means of a Green's function and
a decomposition of the wave function in partial waves.  We have then
found the dependence of the $s$-wave scattering length on the
polarization angle by inspection of the $m=0$ mode, which dominates at
large distances. Equipped with the two-body solution, we have built a
variational wave function of the Jastrow type that has been used as a
guiding function in a DMC simulation of the gas of polarized dipoles
at low densities.  We have found that the scaling of the energy in the
gas parameter is preserved up to values of $x$ where other isotropic
systems deviate significantly. This behavior extends to other relevant
ground state quantities like the pair distribution function, the
static structure factor and the one-body density matrix, including the
condensate fraction which can be determined from the large distance
asymptotic behavior of its isotropic part. 

\vfill

\begin{acknowledgments}

This work has been partially supported by Grants No.~FIS2008-04403
from DGI (Spain), Grant No.~2009-SGR1003 from the Generalitat de
Catalunya (Spain) and Grant No.~P23535 form the Austrian Science
Fund FWF (Austria).

\end{acknowledgments}

\end{document}